\documentclass[12pt]{article}
\usepackage{setspace,curves}
\usepackage{amsmath}
\usepackage{amsfonts}
\usepackage{hyperref}
\usepackage{csquotes}
\usepackage[margin=1in]{geometry}

\newcommand{\R}{\mathbb{R}}

\title{Students' Experience of Cultural Differences Between Mathematics and Physics}
\author{Jeffrey M. Rabin \\ Mathematics Department, UCSD \\ La Jolla CA 92093 \\ jrabin@ucsd.edu
\and
Andrew Meyertholen \\ Physics Department, UCSD \\ La Jolla CA 92093 \\ ameyertholen@physics.ucsd.edu
\and
Brian Shotwell \\ Physics Department, UCSD \\ La Jolla CA 92093 \\ bshotwell@physics.ucsd.edu} 

\begin{document}

\date{\today}
\maketitle

\begin{abstract} How students use mathematics in their physics classes has been studied extensively in the physics education literature. In addition to specific mathematical methods in specific physics contexts, possible effects of more general ``cultural" differences between the two disciplines have also been explored. However, there has been little examination of students' own awareness and interpretation of these differences. We explore the undergraduate student experience of these ``cultural" contrasts, focusing on how they impact learning and problem-solving. Through a qualitative study, including surveys and interviews with students double-majoring in mathematics and physics (or majoring in one and minoring in the other), we investigate students' awareness of distinct pedagogical approaches, mathematical justifications, and organization of concepts in mathematics versus physics classes. We find that students do recognize and navigate these ``cultural" differences, often employing specific coping strategies. We identify specific themes from our data and comment on how students feel that these themes impact their learning. We suggest that increased faculty and student awareness of the identified differences in educational practice could facilitate knowledge transfer between mathematics and physics.
\end{abstract} 

\section{Introduction}

\begin{displayquote} \it
``Philosophy is written in this grand book, the universe, which lies continually open to our gaze. But the book cannot be understood unless one first learns to comprehend the language and read the letters in which it is composed. It is written in the language of mathematics, and its characters are triangles, circles, and other geometric figures, without which it is humanly impossible to comprehend a single word of it."  -- Galileo, The Assayer, 1623
\end{displayquote}

\begin{displayquote} \it
``Before we consider Galileo's demonstrations, it seems necessary to prove how far from the truth are those who wish to prove natural facts by means of mathematical reasoning, among whom, if I am not mistaken, is Galileo ... Therefore anyone who thinks he can prove natural properties with mathematical reasoning is simply demented, for the two sciences are very different."  -- Vincenzo di Grazia, 1613.
\end{displayquote}

Since at least the time of Galileo it has been commonplace to observe that mathematics is the language of science, particularly physics.
Mathematical proficiency is necessary for students to understand physics deeply, and the physics education research literature has extensively investigated how they use their mathematical knowledge in physics~\cite{Caballero_2015}.
Much of this literature is highly specific in terms of both mathematical methods and physical applications, studying for example students' use of integration in electrostatics~\cite{Doughty_2014}, partial derivatives in thermodynamics~\cite{Kustusch_2014}, or linear algebra in quantum mechanics~\cite{Serbin_2022}. 
These studies can be conceptualized in terms of mathematical modeling, transfer of knowledge, and even conceptual blending. 
Certainly, differences in how ``the same" material is presented and used in physics versus mathematics courses can affect students' ability to transfer their knowledge of one to the other. 
But there have also been more general studies of the relationship between physics and mathematics, for example the differences between the ``dialects" of mathematics spoken in each field.
Redish and Kuo (2015)~\cite{Redish_Kuo_2015} describe a number of differences between these dialects, for example that ``loading physical meaning onto symbols does work for physicists and leads to differences in how physicists and mathematicians interpret equations." 
An anecdotal example is ``Corinne's shibboleth" (attributed to Corinne Manogue~\cite{Dray_2002}) which allegedly discriminates well between mathematics and physics faculty: if the function $f(x,y)$ is defined as $x^2+y^2$, then what is $f(r,\theta)$?
Physicists are said to answer $r^2$, while mathematicians answer $r^2 + \theta^2$, showing that they interpret symbols and use function notation quite differently.
This literature suggests ways in which this language barrier may impact students' learning to use mathematics in physics contexts. 
Redish (2021)~\cite{Redish_2021} is an introduction to seven subsequent short articles he wrote to disseminate these ideas to practicing physics teachers, to support their students to ``think about physics with math instead of just calculating." 

Our conception of such ``cultural" differences between mathematics and physics is quite broad, including not only these distinct dialects, but also how and to what extent mathematical claims are justified in each field, how approximations are used, how new concepts are introduced, and differences in pedagogy that may be common to instructors within each field but differ between them. 
All these features go beyond specific topics like integration or electrostatics, but may affect students' ability to ``transfer" or apply their mathematical knowledge in the physics context. 
They form a sort of background to the teaching and learning of individual topics. 

In addition to the work of Redish and collaborators, we have drawn on other papers addressing the cultural aspects of mathematics as applied to physics. 
Uhden et al (2012)~\cite{Uhden2012-UHDMMR} distinguish between the technical skills of mathematics that are used for computation or derivation, and the structural skills that ``are related to the capacity of employing mathematical knowledge for structuring physical situations." 
These authors provide a model for students' use of mathematics in physics that recognizes structural skills in addition to procedural ones and recognizes that physical problems are not simply ``translated into mathematics" as a naive modeling approach would suggest but rather that the math and the physics are inseparable at every stage. 
Palmgren and Rasa (2022)~\cite{Palmgren_2022} give examples of the roles of mathematics in physics drawn from the history of quantum mechanics.
They argue that math is much more than a tool for problem solving, rather ``mathematics contributes to representations of physical information that in turn serve as a basis for further reasoning and modification."

To our knowledge, however, no research has investigated whether students themselves are aware of such ``cultural" differences between physics and mathematics, or of how such differences may affect their own learning. 
We planned our study to include students who were double majors at our home institution, thinking that they would have enough advanced course background in each subject to have experienced a variety of these differences.
To increase the number of participants we also included those majoring in one subject and minoring in the other. 
We distributed a questionnaire to all participants, and after an initial analysis of the responses we selected a subgroup for individual interviews. 
Our questions were quite broad and included the following areas: the role that students think mathematics plays in physics, differences in their experiences in math versus physics courses, stereotypes of mathematicians and physicists, how proof or justification differs in the two subjects, and whether mathematics has been a help or a hindrance in their understanding of physics. 
Based on our own experience and prior research literature, we asked about specific mathematical topics that we expected to differ between the fields or create learning difficulties for students, for example, Fourier transforms, infinitesimals, linear algebra, and the Dirac delta function. 
We expected, for example, that the common use of infinitesimal quantities such as $dx$ or $dW$ in physics would be an obstacle for students in applying calculus as they learned it in mathematics courses. 
We also included Corinne's shibboleth as one question, since we have not seen published data on it. 

Our research questions were the following:
\begin{enumerate}
\item Are undergraduate double majors aware of ``cultural" differences between mathematics and physics, and differences in the presentation and usage of mathematical content between their physics and math courses?
\item What do they cite as major differences in ``common" mathematical content?
\item How do they perceive these differences as affecting their learning?
\item What coping strategies do they employ?
\item What specific classroom experiences do they cite?
\end{enumerate}

Some previous empirical studies have used questionnaires and interviews to try to explore students' experiences along similar lines, but not as broadly as we have. 
de Ata\'ide and Greca (2013)~\cite{Ataide_2012} asked students about the role of mathematics in physics (for example, do you experience difficulty in applying the concept of differential in problems of thermodynamics?), but their results are limited to exploring correlations between students' problem-solving strategies (categorized as Operational Mathematics, Conceptualization, and Mathematical Reasoning) and their epistemic views of math in physics (as Tool, Translator, or Structure). 
de Winter and Airey (2022)~\cite{de_Winter_2022} asked preservice physics teachers in England about the role of mathematics in physics at both university and school level.
Responses did show awareness of different treatments of ``the same"  subjects in math versus physics courses, but were vague as to examples. 
Kapucu (2014)~\cite{Kapucu_2014} asked Turkish preservice elementary math/science teachers about their conceptions of mathematics, physics, their relationship, and the learning of each, but it appears that their participants had only majored in these subjects in high school. 
Responses were vague (for example, ``physics is primarily based on calculations, so requires math") and only coarsely classified as ``fragmented" versus ``cohesive" and ``lower level" versus ``higher level".  
Schermerhorn et al (2019)~\cite{Schermerhorn_2019} were narrowly focused on students in a spins-first quantum mechanics course at three universities and asked whether the mathematics helped students understand the physics concepts, whether the reverse was true, and which was more challenging. 
Some student comments resonate with those of our subjects, for example, ``The physics makes the math easier to visualize and the math provides a base for physical intuition. For example, the properties of wave functions helped me understand Fourier transforms" (ibid., page 534).

\section{Design of Study}

We strived to make the survey questions unbiased --- we did not want to show any preferences for math vs.~physics classes or instructors, and we tried to formulate questions that would allow students to discuss pros and cons of their experiences with both subjects. After agreeing on the survey questions,\footnote{We give the survey questions and summaries of responses in the next section.} the authors sent out invitations to all math/physics double majors at our institution, and also to those students majoring in one subject and minoring in the other. We sent out 36 emails, offering each student a gift card to complete the $\sim\!1\text{hr}$ survey. 24 students responded and were interested in participating in the study.

At this point, we numbered students roughly in the order they responded (numbered 1-24). In addition, the students were given a letter code to indicate their major combination:  ``D" for double majors, ``M" for math majors with physics minors, and ``P" for physics majors with math minors.\footnote{Although we did not select on the basis of class standing, all 20 students who completed the survey had JR/SR standing, and all 11 interviewees had SR standing. These interviewees were a mixture of 3rd-year students (who had enough credits for SR standing) and 4th-year students nearing graduation.}

The authors were able to collect 20 completed written responses to the surveys. After analyzing the surveys individually, we met to discuss the responses via a thematic analysis~\cite{Terry_2017} of the surveys, flagging significant quotes and determining potential emergent themes.
We extended an interview invitation to those 13 students whose responses included experiences and opinions that we wanted to follow up on. Of these 13 students, 11 agreed to participate in an interview with one of the authors (with audio recorded).
Interviewed students were compensated with another gift card. 

Interviews were designed~\cite{Knott_2022} to be semi-structured, based on a general framework of questions that we asked each student, but we also followed up on specific statements made during the interview. Also, we followed up on statements made in the student's survey responses. If a student mentioned a specific mathematical topic in their written survey (see Section~\ref{survey-questions}, Question 6), then we asked them in the interview to describe their learning experience in math and physics classes involving the topic. We split up the students randomly among the three authors and each author interviewed 3-4 students.

After completing the interviews, the authors transcribed each of their interviews using the audio recording. We created and shared summary notes and memos. We then met to discuss each interview in depth, highlighting quotes that stood out to us, or ideas that seemed to come up multiple times.\footnote{We also reviewed the survey answers after the interviews; we didn't find anything mentioned in the ``non-interviewee" survey responses that was not also represented in the interviews.} After this, we were able to categorize the main results into four principal themes, which we discuss in detail in Sections~\ref{theme-1}-\ref{theme-4}.

The following table summarizes students' participation at various stages of the study:

\vspace{0.25cm}

\begin{center}
\noindent
\begin{tabular}{|l|c|c|c||c|}
\hline {Stage} & { \#D } & { \#M } & { \#P } & Total Students\\
\hline
\hline Request for participation & 13 & 8 & 15 & {\bf 36} \\
\hline Surveys sent out via email & 10 & 3 & 11 & {\bf 24} \\
\hline Surveys received/analyzed & 8 & 2 & 10 & {\bf 20} \\
\hline Interview invitations extended & 4 & 2 & 7 & {\bf 13} \\
\hline Interviews conducted/analyzed & 4 & 1 & 6 & {\bf 11} \\
\hline
\end{tabular}
\end{center}

\vspace{0.5cm}

\section{Survey Questions and Results} \label{survey-questions}

The survey consisted of the following questions. A quick summary of the results directly follows each question.

There were no questions where there was a clear distinction of answers by major (other than question \#1, of course). That is, there were no questions for which double majors responded significantly differently from physics majors with a math minor (or from math majors with a physics minor). However, our sample size is probably too small for this to be meaningful.

\begin{enumerate}
\item {\bf What is your major (and minor, if any)?}

(All students were double majors or one major and one minor in math and physics, by design.)

\item {\bf If you chose one major/minor first, and added the second later, what was the reason you chose to add the second?}

Most students said that this was done out of interest and/or that the minor subject complements the major subject.

\item {\bf Suppose that $f(x,y)=x^2+y^2$. What is $f(r,\theta)$? Do you think that either a mathematician or a physicist might give a different answer than you did? What answer, and why?}

Only one student (D4) was able to identify that the answer \emph{could} be $r^2+\theta^2$ (the student provided both the ``physicist" and the ``mathematician" response). 17 others only gave the physicist's response, that the answer is $r^2$. Two students said $f(r,\theta) = r$, with one adding, ``A physicist would probably not be concerned with the square on the radius."

\item {\bf Have you been dissatisfied with an explanation, derivation, or proof given in a physics class because you felt it was not mathematically precise or rigorous enough? Can you give examples?}

15 out of 20 students cited at least some dissatisfaction, especially with Fourier analysis, Hilbert spaces, and differential equations. In addition, students did not enjoy the ``hand-waviness" of mathematical manipulations and/or approximations made in physics classes. Others were satisfied with physics classes not spending too much time on the math.

\item {\bf Have you had difficulty understanding mathematical techniques presented in a physics class because the presentation or notation differed from what you had seen in previous math classes? Can you give examples?}

A great majority of students either said ``no" or cited the difference in which variable was used to represent the polar angle in spherical coordinates ($\theta$ vs. $\phi$) in physics vs. math courses. Nearly all of the students who cited the polar angle notational difference said that they got used to this.

\item {\bf Have you found any of the following specific topics challenging to learn about or use because of differences in how they are presented in math versus physics? Conversely, are there instances in which your understanding of a topic was improved by having both a physics and a math perspective on it? \label{topics-survey-questions}}
\begin{enumerate}
\item {\bf The Dirac delta function (quantum mechanics, electromagnetism)?}
\item {\bf Infinitesimal quantities or vectors such as dx and $\vec{\bf ds}$, used extensively in physics but not so much in math?}
\item {\bf Concepts and terminology from linear algebra and matrix theory, such as basis, linear operator, eigenvalues?}
\item {\bf Calculations involving derivatives or integrals of functions?}
\item {\bf Calculations involving infinite series?}
\item {\bf Methods for solving or understanding differential equations?}
\item {\bf Fourier series and Fourier transforms?}
\item {\bf Dimensions or units of physical quantities, emphasized in physics but often ignored in math?}
\item {\bf 3D vector calculus (div/grad/curl, Stokes’ theorem, etc.)?}
\end{enumerate}

Students cited the Dirac delta function and Fourier analysis as topics they never saw presented in math classes; they had some difficulties with the topics as they were presented in physics. Several also mentioned that infinitesimals were not treated rigorously in math courses, but that they were used enough in physics that it would have been beneficial to have seen them more in math. For the most part, students said their understanding was improved by having seen 3D vector calculus and differential equations in both subjects, and nearly everyone agreed that their ability to calculate derivatives/integrals was improved by exposure in both subjects. Responses about linear algebra were mixed, with some saying that the math exposure was helpful for upper-division quantum mechanics, while others said that their math exposure was not sufficient. Not much was said about infinite series or dimensions/units.

\item {\bf Have the different standards of proof/explanation/justification/evidence accepted in physics versus math classes had effects (positive or negative) on your learning, or your understanding of the material, in those classes?}

(Numbers below indicate the number of responses in each category.) 

(7) Yes: proofs and justifications are more rigorous in math classes than in physics classes, which has had a negative effect on my understanding in physics classes due to lack of clarity; (6) Yes: the two complement one another nicely, which has had a positive effect on my learning; (5) No: I recognize there are differences, but I don't think this has affected my learning overall. Other answers were either not clear or a blend of these.

\item {\bf Along with the definitions, equations, laws, theorems, and so forth taught in physics and math courses, instructors and textbooks also try to convey intuitions about these topics (“how you should think about them or visualize them”) that are appropriate to each subject. Have you noticed differences between the intuitions taught or expected in math versus physics, and have any of these differences had effects (positive or negative) on your learning?}

A few students said that they didn't notice a difference, but most said that intuition is more emphasized in physics classes. Students varied in which they preferred (math providing a better framework, but physics providing more concrete examples to visualize). A few mentioned that math and physics have different patterns of reasoning: math classes start fully general and only later provide examples or specific applications, while physics classes start with specific cases to build intuition, and then try to generalize.

\item {\bf Have you experienced explicit comments from either physics or mathematics instructors that are critical of how some topics are presented in the other discipline? Can you give examples?}

(8) Minor or only playful joking or minor comments; (12) No. Most examples involved physics instructors saying something to the effect of ``mathematicians wouldn't like me doing this, but..." (not explicitly critical of the other discipline).

\item {\bf Have you asked questions of your mathematics or physics instructors about topics from the other field, and been unsatisfied with their responses? Can you give examples?}

(0) Yes, from math instructors; (7) Yes, from physics instructors; (13) No. Topics where the physics professor's response was unsatisfying: math behind special relativity (for example, Lorentz/Poincar\'{e} groups), various group and set theory concepts, expectation of an operator, special cases of integrals, bracket notation in quantum mechanics or probability, divergence and curl in electromagnetism.

\item {\bf What suggestions would you give to physics or mathematics faculty about how to help students better understand material that overlaps or is common to both subject areas?}

Suggestions to math professors: give real-world and physics applications.\\
Suggestions to physics professors: be more organized with the presentation of material, be more clear/precise/rigorous with the math.\\
Suggestions to both: be clear about differences in notation and conventions (for example, factors of $2\pi$ in Fourier transforms), provide resources to students to better understand math/physics connections.

\item {\bf When the presentation of a topic in a lecture or a textbook is different between your math and physics classes, do you have strategies for reconciling them or coping with the differences?} 

(6) Use online resources; (5) Try to understand the differences; (5) Ask a professor; (3) Keep them separate; (3) Hasn’t been an issue;
(2) Give the math understanding priority.

\end{enumerate}

\section{Summary of findings in specific math topic areas}

Following is a brief summary of student comments about specific mathematical subject areas that we asked about. We will revisit many of them in the subsequent discussion of our major themes, providing interview quotes for support. 

\subsection{Infinitesimals}

We anticipated that the common use of infinitesimal quantities in physics would conflict with students' prior understandings of calculus from their mathematics courses. Physicists happily work with infinitesimal amounts of charge $dq$ or displacement $dx$ whereas mathematics courses treat calculus using limits and often explicitly deny that infinitesimal or differential quantities are individually meaningful. However, our students did not identify this as a major obstacle. Some of them recognized physics infinitesimals as an indirect way to reason about limits or small finite increments, which could be rigorously reformulated in those terms if desired. Although they were not bothered by the supposed {\it infinitesimal} nature of these quantities, they did express some confusion over the rules for {\it manipulating} them, which they included in their catalog of mysterious physics ``tricks". For example, one student realized that canceling two infinitesimals would solve their homework problem, but struggled to justify this manipulation. Others were unsure when differentials could be separated from derivatives and manipulated independently, or when it was permissible to change partial derivatives to total derivatives and vice-versa. Cancellations that look intuitive in single-variable calculus, for example in the Chain Rule, appear less so when partial derivatives are involved. 

\begin{displayquote} \it
``I know that if I cancel them then [infinitesimals], then I would be able to get an answer pretty easily, but there's a reason that I was stuck on this problem for an hour and it wasn't that I didn't see that I could cancel them, it was that I kept looking at it and saying, well {\it why} can I do that?" (P1) 
\end{displayquote}

\subsection{Fourier Analysis}

Another topic that we expected the students to have observations on was Fourier analysis. This subject is not generally discussed in the introductory calculus/linear algebra/differential equations sequence but is crucial to an introductory physics curriculum. Fourier analysis may be encountered later in (non-required) mathematics courses such as partial differential equations or numerical methods. Our physics majors are introduced to the Fourier transform in the optics portion of the advanced first-year physics sequence but they find the treatment of it there too brief. Some were unfamiliar with the complex exponential function $e^{ikx}$ and cited the differences between this notation and the alternative sines and cosines as a challenge.  Some pointed out that the Fourier transform is defined as an improper integral but is rarely evaluated directly from that definition. Rather, ``tricks" such as completing the square are employed. Many students mentioned their physics math methods course as being the first place they were given enough time to fully learn and appreciate Fourier analysis. 

\subsection{Delta functions}

Another topic we specifically asked students about was the Dirac delta function. Like Fourier analysis the delta function is not discussed in the first-year math sequences (or perhaps anywhere in their math coursework). Most students were willing to accept it in the context of physics applications despite feeling that they did not understand it deeply as a mathematical object. They knew the rules for manipulating it and could do so despite wondering how these were justified. One student wondered how $a \delta(x)$ can differ from $\delta(x)$ if both are infinite at $x=0$ and zero elsewhere. A mathematics major who had seen the delta function in quantum mechanics took the view that since quantum mechanics violates our macroscopic intuitions anyway, it makes sense that its mathematical tools such as the delta function would also violate them. Several students mentioned the integral form of the delta function, $\delta(x) = \frac{1}{2\pi}\int^\infty_{-\infty}dk \, e^{ikx}$, specifically as something that seemed ``like magic." 

\subsection{Corinne’s shibboleth}

Dray \& Manogue (2002)~\cite{Dray_2002} propose that the following question will distinguish physicists from mathematicians: \\[-5pt]

\fbox{
\begin{minipage}{0.85\textwidth}
One of your colleagues is measuring the temperature of a plate of metal placed above an outlet pipe that emits cool air. The result can be well described in Cartesian coordinates by the function
\begin{center}
$T(x, y)=k(x^2+y^2)$
\end{center}
where $k$ is a constant. If you were asked to give the following function,\\
what would you write?
\begin{center}
$T(r, \theta)= \ ?$
\end{center}
\end{minipage}
}

\vspace{0.25cm}

The physicist, considering $T(x,y)$ to be a function representing the physical temperature at a point in space, would infer that $T(r,\theta)$ is the same temperature function but expressed in 2D polar coordinates instead of 2D Cartesian coordinates. The mathematician, instead, would consider the function $T: \R^2 \to \R$ as a mathematical object independent of the dummy variables used to represent the two real number inputs. As a result, the physicist will answer ``$T(r,\theta) = kr^2$\,", and the mathematican will answer ``$T(r,\theta) = k(r^2+\theta^2)$".

In our version of the prompt, below, we omitted the physical context for the function, hypothesizing that this would remove potential bias toward the ``physicist" response.  In addition, we explicitly suggested that there might be other possible answers given by a physicist or mathematician, to encourage students to apply multiple perspectives to the problem: \\[-5pt]

\fbox{
\begin{minipage}{0.85\textwidth}
Suppose that $f(x,y)=x^2+y^2$. What is $f(r,\theta)$? Do you think that either a mathematician or a physicist might give a different answer than you did? What answer, and why?
\end{minipage}
}

\vspace{0.25cm}

\noindent Surprisingly, \emph{all} students gave the physicist's response (or a variant of it\footnote{Two students said that $f(r,\theta) = r$, which we consider closer to the physicist's answer than the mathematician's answer.}), and only one recognized that a mathematician might possibly give a different answer because of an assumption made in the physicist's response:

\begin{displayquote} \it
``It’s possible someone who doesn’t study physics would be confused by/not understand the abuse of notation implied here, where the actual function f is a different mapping depending on what symbols (coordinates) we use. They might instead treat the question “What is $f(r,\theta)$” as a request to directly substitute the symbols “$r$” and “$\theta$” into the original function." (D4)
\end{displayquote}

Student D4 mentioned that the mathematician's response is the more ``rigorous" one and clarified that in his view the physics interpretation is the abuse of notation. Also, this student mentioned that they had thought about this issue before via discussions with their math-major close friend.

Redish and Kuo~\cite{Redish_Kuo_2015} identify that physicists load meaning onto variables in a way that mathematicians normally do not. We confirmed a strong version of this in our study with our version of Corinne's shibboleth --- someone giving the physicist's response is clearly assigning meaning to the coordinates $(x,y)$ and $(r,\theta)$, even though the identification of these variables as Cartesian and polar coordinates was never given in the problem statement. In the interviews, we tried to probe the students' answers a bit more, asking if their answer would have changed if we instead asked about the function $f(u,v) = u^2 + v^2$ or $f(p,q) = p^2 + q^2$ (and then asked what $f(r,\theta)$ would be in that case). It seemed the interviewees had already decided the question had to do with changing coordinates, and still gave the answer of $f(r,\theta) = r^2$. Either this, or they said that they wouldn't know how to answer without knowing the meaning of the new variables. 

We do not find it surprising that undergraduate students associate definite meanings with variables in this way. What is surprising is that even mathematics majors do not suggest the mathematician's interpretation of function notation when prompted for an alternative.

\subsection{Role of Math in Physics}

We asked our interviewees for their views on the role that mathematics plays within physics. The most common response was that mathematics is a tool for solving problems and understanding the physical world. The term ``tool" includes the implication that physicists do not always need to delve into or justify how the tool works; for their purposes it is enough that it does work and the underlying reasons can be left to mathematicians. A related view, echoing Galileo, was that math is a language for describing the world. Physics needs a language that is more precise and more adapted to logical reasoning and calculation than English, and mathematics fills this need.

More detailed responses suggested that mathematics provides a general framework that physicists can then fill in with boundary conditions, physical parameters, or other ``limitations" that describe the specific laws of physics and material properties applicable to a given problem.

\begin{displayquote} \it
 ``Physics is just applied math with concrete, like, limitations... like oh, the ball is not going to go outside of the room... we’ll never, when doing physics, ... you’ll never consider something outside of a third dimensional realm." (P20)
\end{displayquote}

\begin{displayquote} \it
 ``In cosmology, the Einstein equations and the Friedmann equations are all just math, the physics is when you then say, okay, well this is the, this is our idea of what the dark matter content is, and this is our idea of the equation of state of the matter, and we plug that into the stuff that’s been mathed out for us and then we get, you know, we get the scale factor throughout the expansion history of the universe." (P1)
\end{displayquote}

\section{Major Theme 1: the nature of the subjects} \label{theme-1}

A major theme in our data is that students perceive mathematics to be deductive, standardized, and highly structured, whereas physics is inductive, eclectic, and flexible. 
These observations apply to the nature of proofs or arguments presented in each subject, the presentation of other material in the classroom and in textbooks, the homework and exam problems posed in each class, and the ways in which instructors respond to questions from students. 
The presentation of mathematics is often structured by the Definition-Theorem-Proof format, even when ``Motivation" or ``Application" are added to this template. 
Physics is more likely to proceed inductively from examples of phenomena to generalization and theory. 
When a physics class requires a new mathematical concept or technique it is likely to be presented briefly in the form of an equation or rules and then applied without a deep justification or motivation. 
Students recognize that mathematics is abstract and not necessarily connected to the physical world, whereas physics must be so connected even at its most abstract. 
Physics is about explaining or predicting real phenomena, while mathematics is ``about" itself. 
In that sense physics is a blend of theory and experiment, while mathematics is ``all theory". 

\subsection{The Nature of Proofs/Arguments in Each Subject}

Students perceive mathematical proofs as highly structured, with the hypotheses and conclusion announced at the start and all the logic spelled out explicitly. 
Specific named methods such as Mathematical Induction may form the proof structure, and named theorems may be quoted in support of particular steps. 
A physics ``proof" is more free-form and can contain jumps and gaps; it looks more like a calculation than a piece of reasoning, and the result may appear suddenly as a surprise. 
(One student gave the example of the speed of light emerging suddenly from a calculation based on Maxwell's equations.) 
The steps typically follow by standard computational tools from algebra and calculus without further overt justification. 
Mathematical proofs are more wordy, the words serving to explain the reasoning.
In contrast, when words appear in a physics proof they may interrupt the flow of the calculation/proof to introduce a physically motivated approximation, which does not have a rigorous justification, or perhaps to dismiss a case as ``unphysical" rather than mathematically disproved. 
Unlike a mathematical proof, a physics proof may not be self-contained, and students feel that they have to ask additional questions after class if they want to understand the transitions between steps. 
Students also mentioned that mathematical proofs may be more rigid in terms of how terminology or symbols must be used, or how the logic must be sequenced, while physics proofs are more flexible. 

\begin{displayquote} \it
``Once I go through a mathematical proof, ... as long as I understand the intermediate steps, I end up with, like, I know why this result works inside and out. And, whereas physics proofs sometimes require a little bit more, in some ways, trust." (D22)
\end{displayquote}

\begin{displayquote} \it
``Whereas in physics sometimes if you look at the notes for your proofs there's not -- they have to make a jump but I'm not entirely sure how they got there, then I have to go talk to them, like how do you get from here to here?" (M9)
\end{displayquote}

\begin{displayquote} \it
``Proofs in math class are more like a written paragraph type of thing versus in physics ... it is all just math like algebra and calculus ...  and a lot of the proof in physics is in the math rather than in the spoken words." (P20) 
\end{displayquote}

These comments are consistent with the view of a student in ~\cite{Redish_Kuo_2015} , who said (page 579), ``[In math] it just all makes sense to me, because there's a reason it works, and it's just one reason. It's not like in physics really where there's so many different cases like I said before. In math, if I understand the proofs of why it's that way, and then, I'm comfortable using that equation."

\vskip 10pt

\subsection{How Topics are Presented in Class}

Students felt that topics in mathematics classes are developed thoroughly through the definition/theorem/proof structure, so that their properties are built up gradually and fully justified. 
In physics classes math is a means to an end, and new techniques may be simply presented and then exemplified through applications. 
Students recognized that the main focus is on the physics, while the math is simply a tool which can be accepted without rigorous justification if it works.
The rules for calculating with it take precedence over complete definitions or derivations. 

\begin{displayquote} \it
``My math classes have been, they have, like a familiar structure ... they have some motivation, they have some definitions, a theorem, and a proof. ... Whereas in physics I feel like it's kind if been more all over the place. Like the transitions between when somebody's giving like a heuristic argument or an actual proof of what they're trying to show, it's a bit blurry." (D3)
\end{displayquote}

\begin{displayquote} \it
``It's also a lot easier to take notes in a math class for that reason. ... I can see what each thing is defined as, what the main theorem is, and how it's proved. Whereas in physics I think things are a bit at least, in my experience, a bit more, like haphazard." (D3)
\end{displayquote}

\begin{displayquote} \it
``In physics the words that you would be using in that are generally the physical explanation of the phenomena where you are kind of saying like this is roughly how the equation matches up to the phenomena, while in math you're not trying necessarily to match up the equation to a phenomenon, you're simply trying to explain the equation and how it works." (M9)
\end{displayquote}

\begin{displayquote} \it
``When they present you math in physics they don't tend to present words alongside it." (M9) 
\end{displayquote}

\begin{displayquote} \it
``[Math topics are presented as definitions, theorems, proofs, but] in the physics class I feel like often we're given an equation maybe and then we jump right into uses of it, like we're given the Fourier series and then we jump right into applying it." (P1) 
\end{displayquote}

\begin{displayquote} \it
``In a physics class math is presented as a means to an end rather than something to study itself ... one example is in classical mechanics we kind of did a crash course into calculus of variations for Lagrangian mechanics." (P20) 
\end{displayquote}

\subsection{Nature of Homework and Exams in Each Subject}

We will have more to say about assignments in each subject when we discuss student perceptions of ``tricks" in physics. 
However, students find math assignments more predictable than those in physics, particularly the expectations for what correct or complete solutions should include.
The stricter standards of justification in mathematics and the presence of physically motivated approximations or shortcuts in physics contribute to this. 

\begin{displayquote} \it
``If I have my math homework and my physics homework on my desk, and someone comes by and looks at it [and asks] `Oh, what class is this?' I'm like, ... if it looks like an essay, then it's math. If it's math, then it's physics." (D5)
\end{displayquote}

\begin{displayquote} \it
``I can get full credit on the assignment even though I don't know why the thing I just applied works in this instance ... you can BS the physics assignment and still get 100 on it, even though you don't understand why the stuff you're writing works ... I can explain the steps and still not understand why they work, versus in a math problem, in a math homework assignment, if you BS the homework assignment you're not getting 100 on it ... because you're going to be missing some crucial steps in your proof, and you know you're missing those crucial steps, and when a proof is complete you feel it." (P1)
\end{displayquote}

\begin{displayquote} \it
``The midterms and the finals in math classes are so much more standardized, like I feel like I know exactly what I'm getting into in a math exam even if it's with a whole new professor. Like with all the math exams of the [calculus] series, they kind of felt the exact same, and then now that I took a couple proof-based math classes, those kind of exams all felt the exact same too. Versus the physics exams it kind of feels like there's like a massive variance, like a swing between them ... my grades in math exams are very consistent like I feel like I get pretty much the same score on every math exam even in different classes, versus in physics classes, like I've gotten anything from above a 100 to below a 20 in physics classes and it's not because I'm studying any differently, just because the exams vary so much." (P1) 
\end{displayquote}

\subsection{Responses to Student Questions}

Some students felt that their mathematics professors were more likely to give complete or satisfactory answers to ``why" questions.
Physics professors might not know the answers to technical mathematical questions or might de-emphasize the importance of such issues relative to developing physical understanding. 

\begin{displayquote} \it
``In physics, generally, ...  we use it because it works, not because it's like, we never, like, prove anything ... If I ask some question, or if I hear somebody else ask a question, like related to the math, ... like `How are we able to use this? What does it mean?' If it's, like too math, like `jargony', in general the professor seemed to be like, `It doesn't really matter, like don't think about it ... you just have to know that it works.'" (P11) 
\end{displayquote}

\section{Major Theme 2: physics ``tricks"} \label{theme-2}

Another major theme from our data focuses on categorizing ways in which mathematics, as it's used in physics classes, can appear unclear, sloppy, handwavy, etc. In general, we call such sources of confusion physics ``tricks."

Physics classes can gloss over mathematical details. Often, this is due to lack of lecture time, with physics instructors choosing to spend their limited time focusing on physical phenomena rather than mathematical manipulations. Sometimes students are fine with this, but in other cases the mathematical manipulations performed (or omitted) in a physics class are a big source of confusion. Such sources of confusion fall into two categories: “techniques” and “approximations.” Roughly speaking, the former are mathematical objects, structures, or manipulations that can be mathematically justified or further explained, although sometimes this is not clear to students and may be omitted by their instructors. It is mostly students’ unfamiliarity with the techniques or mathematical subject matter that leads to their confusion. The latter, “approximations,” involves approximating exact expressions, “erasing” information, and reducing general expressions to specific cases, sometimes as a “shortcut” to get around difficult math. These often require physical arguments specific to the physics problem at hand.

It is worth noting that students’ lack of conceptual understanding of the material, and lack of intuition about what the math is representing, can contribute to their confusion for both types of “tricks.” Also, we note that students became more comfortable with the use of many ``tricks" over time, after initial discomfort when first encountering them.

\subsection{Mathematical Techniques}

Some of the mathematical “tricks” that students mentioned can be classified as mathematical techniques, manipulations, or entirely new topics. Generally, these are things that could be encountered in a math class, as they don’t rely on physical arguments or approximations.

For the most part, students were okay with an abuse of notation or somewhat “sloppy” mathematical methods, so long as they had some exposure to the mathematical topic. For example, the authors anticipated more complaints about the interpretation of differentials in physics classes (e.g., treating $dq$ in electromagnetism as a small amount of charge), but largely this was okay with students. A few students said that they would have liked to have learned a bit more about differentials in math courses, but that they picked up what they needed in physics (especially upper-division electromagnetism). Another topic was the order of taking integrals or derivatives – a few students who had a bit more math background mentioned that physics instructors tended not to check the conditions justifying the interchange of the order of integrals/derivatives/summations, but that largely this didn’t affect their ability to follow along in the courses. When asked a followup question about some steps of proofs not being justified in physics classes, one double-major said:

\begin{displayquote} \it
``Uhh, they went like slow enough, whereas, like, if they did something, I could just kind of see why it worked out. Like, even if he/she didn't justify it like, I can just like, think, `Oh, this, works,' so I didn't really... there's nothing that really stood out to me." (D3)
\end{displayquote}

One topic that confused several interviewees was Fourier analysis (Fourier series and transforms):\footnote{Most of the interviewed students took a 5-quarter honors sequence of introductory physics, which goes into a little more detail than the standard sequence intended for engineering majors. In this honors sequence, Fourier transforms are brought up in either the 4th or 5th quarter, first with optics and later with quantum mechanics.}

\begin{displayquote} \it
``I would say the Fourier transform... It was probably, I mean, it's definitely like the most foreign piece of math." (D5)
\end{displayquote}

There seemed to be a lot of factors contributing to this: the topic was new when they were introduced to it in physics, it was introduced for a specific purpose (to discuss optics) and not discussed in full generality, and homework exercises tended to focus on particular manipulations that were important in performing calculations, but didn’t focus on conceptual understanding of Fourier analysis in general. It was these calculations in particular that were almost uniformly called “tricks” by the students --- they didn’t know why they were doing what they were doing, didn’t know the “rules,” in effect, and didn’t know what “tricks” were allowed or justified. This was especially pronounced when the students reflected on their homework assignments in these courses:

\begin{displayquote} \it
``The homework for [upper-division math methods course in the physics department] was, was really complex and, and very difficult and it required some creative manipulations of the Fourier transform and of complex numbers that like, like we had to manipulate it in interesting ways, in order to do that you need to understand exactly what the, the Fourier transform is and what these operations are doing, and without really understanding exactly what they're doing it, it just felt like these problems were kind of magic and it's like, oh look, there's this trick to solve it but you know tricks aren't helpful if you don't understand the operation that they’re, that they're getting around." (P1)
\end{displayquote}

\begin{displayquote} \it
``I personally felt like there was a lot of like tricks that are used in the Fourier transforms that are just kind of well-known and you just have to look up. But I remember like I didn't like know how to like… I didn't know they were tricks, so I would like try to actually solve them. And I it would take me so long, and I would like get together with my friends, try to like, solve these problems, and it would take me like so long. And then we just look it up, and then it's actually quite easy and it's just like a well-known fact, or whatever. So that was kind of tough." (P11)
\end{displayquote}

A similar sentiment was expressed at the upper-division level with linear algebra in quantum mechanics. Almost uniformly, students mentioned that the one quarter of lower-division linear algebra that they had in a math class was insufficient preparation for upper-division quantum mechanics. The lower-division course never focused on abstract (or infinite-dimensional) vector spaces, nor used the field of complex numbers, nor discussed hermitian/unitary operators. As a result, there was a similar feeling of being overwhelmed in quantum mechanics as with Fourier analysis, as the students were trying to learn the math along with the particular physics applications, with new notation on top of it all:

\begin{displayquote} \it
``...in linear algebra you have a matrix and eigenvalues and  corresponding eigenvectors but I feel like that sort of calculation, or like I guess that calculation algorithm type thing that you would do in linear algebra is not at all how it was done in quantum mechanics. And that's definitely more of a notation thing in physics where you use the Dirac notation, you know, like operators and all that stuff, you hear a lot about like, and then so it's like that what steps in the calculation are done I feel like in physics they assumed like, `oh they learned this in, like linear algebra,' but obviously we didn't learn Dirac notation in linear algebra." (P20)
\end{displayquote}

There were a few other mathematical topics mentioned, but not as often as the above two. These include symmetry arguments (for evaluating integrals or determining the direction of vector fields) and solutions to differential equations. The authors suspect that these issues were not as universally cited because the students in general knew the physical context and/or they had some familiarity with the mathematical techniques required.

\subsection{Approximations}

Students brought up approximations in a few different contexts, with Taylor series being the most common. In general, students are uncomfortable continuing with an approximate expression if they’re aware (or believe) that an exact expression exists, if they don’t see how the approximation will help solve the problem at hand, or if they don’t think they’re developing tools/skills that will generalize and help them tackle other problems on their own. A related issue was mathematical “shortcuts” taken during lecture: going from general expressions to specific cases, which is a different kind of erasing information.

For example, regarding back-of-the-envelope calculations, one student expressed discomfort when a professor would plug in numbers, but not their exact values. An example might be canceling out $\pi$ in the numerator with $3$ in the denominator of an expression when an instructor is estimating numerical values. Students felt a disconnect between doing “real” (“precise” or “exact,” in their eyes), difficult physics problems, and being cavalier when getting a final result:

\begin{displayquote} \it
``[The professor] always liked to mention, like back-of-the-envelope calculations. And [laughs] I remember a lot of people really were kind of upset by that. He/she liked to just approximate a lot of stuff very quickly, and I remember people were kind of like... people didn't really like that at first... You're like, `why can't we just, why don't we just use like, the most exact answer?' But then later on, I realize it's like, it's really like, negligible. It's like there's no point in including all the extra stuff." (P20)
\end{displayquote}

Many students mentioned Taylor series when the topic of approximations came up. One student, a double major who had already taken several advanced mathematics courses, gave the following ancedote:

\begin{displayquote} \it
“An example that I've thought about where it's like, physics kind of like sweeps a bit of stuff under the rug. I think for like, you know like the pendulum swinging? Umm, they make an approximation where the angle is small. So normally when like you Taylor expand it you have a bunch of trig terms, but they just do like $\sin \theta$ is approximately $\theta$, and it seems like, you know, as somebody coming from math, there's a whole lot of other information in those Taylor expansion terms… hypothetically, if I were to ask a professor like, `What do you do about those other terms, like isn't that part of the physics?' And they say like `no, it only matters that the angle is small, and we can make this approximation.' It's like, well that's not satisfying to somebody like, likes the full solution, I guess." (D3)
\end{displayquote}

This student actually mentioned that their math background was a hindrance in physics problems – they are not used to discarding information in a seemingly arbitrary manner. What are the rules behind when students are allowed to linearize equations? When should we keep the second-order terms? Why are we not trying to quantify the error when throwing away these terms? Shouldn’t the error accumulate over time in a problem like the pendulum? And, if so, why are we ignoring this?

When asked later if there was an exact answer to that problem, the student replied:

\begin{displayquote} \it
“I think, well I'm pretty sure there is an exact answer. I think I knew that, because like the differential equation isn't that hard, like $x'' = \sin x$.” (D3)
\end{displayquote}

\noindent This confusion is probably not too surprising, as a 5-10 second quick remark in lecture to the effect of “this equation is not solvable analytically” or “a complete solution involves elliptic integrals” can easily be missed. Instructors may say that advanced physics classes cover things in more detail, implicitly promising that a more exact treatment exists and awaits them in a more advanced course. Understandably, students might imagine that we’re always “sweeping things under the rug.”

Another student mentioned that if such approximations were to be made in a math class, then there would be more rigorous analysis of error bounds:

\begin{displayquote} \it
``I've taken real analysis, when you do real analysis and you're giving the approximation functions for equations in that class like the Taylor expansions, they of course give you the error and they prove why the error works and they are like, this is how we minimize the error to make sure that there is as little error as possible. We weren’t given anything like that [in the physics course]... I personally find it a little bit confusing to have that stuff and then not have any explanation for why these work the way they do and like what the margin of error is and how we can minimize that margin of error, especially since it feels like it would be important." (M9)
\end{displayquote}

When students mentioned Taylor expansions in interviews, they cited negative experiences of homework problems where they were told explicitly to Taylor expand some expression to a certain order. Many said that Taylor series appeared often and seemed important, and that they wished they had more introduction to it in their classes. However, only one gave the sentiment that Taylor series/expansions can be a useful tool to analyze physics problems.

\section{Major Theme 3: when are preparatory math courses helpful?} \label{theme-3}

Our next theme centers around how physics students perceive their preparatory math courses. In general, the student feedback that we collected seems to support the claim that these math courses are always at least somewhat helpful. However, some appear to be more helpful than others. 

For most physics majors, their first college math courses are the introductory calculus sequence. At our institution, this sequence is a series of five one-quarter courses. Based on our interviews, students taking introductory physics seem to universally appreciate introductory calculus and feel it prepares them well for introductory physics. For example:

\begin{displayquote} \it
``And so learning, being able to learn, like, something in [vector calculus] like a path integral or a surface integral, and then like a week and a half later it comes up for the first time in [electromagnetism], it made it a lot easier, or it’s like, for one, it’s your second time, like our professor did a good job of explaining like, oh, this is what a path integral is, that's like a review thing it helped to hear it a second time in both contexts like that and then seeing it applied to a physical system and so that helped, so then obviously learning it first in math made it easier to apply in physics but seeing a very concrete physical solution and in a way having a whole class dedicated to concrete physical solutions made [vector calculus] a bit easier to kind of digest.” (P20)
\end{displayquote}

In the quote above, vector calculus seems to be especially helpful for electricity and magnetism, as one might expect. In math classes in particular, there does seem to be a tension between physical application and theoretical exposition (often roughly measured by the time spent introducing and developing proofs.) Whereas the introductory calculus series seems primarily interested in applications and building mathematical skills, a good math program is invested in fostering proof writing ability and thus some courses end up spending time in this pursuit. A sub-theme that we noted was how the balance of time spent by a math instructor on physical applications vs. more pure mathematical theory affected student learning. There seems to be student consensus that a more ``proofy" course is less helpful when it comes to the learning of physics. For example, we have a quote comparing the complex analysis and the ``Introduction to Proof" courses in the mathematics department:

\begin{displayquote} \it
``Like when you multiply, the angle just kind of goes around. And when you divide, the angle goes back the other way. That kind of stuff. When you multiply, it just increases. And then, the visualization of umm, the, what was it, the branch cut and stuff, where you know, it becomes like a spiral, and it keeps going around. I was like, ``Wow, this is, this is really cool.” It was just really interesting to me. But then [Intro to Proof] was like a lot of these symbols, symbols. You have to write a proof in this way, this way, and I was like, it wasn’t as interesting.” (P12)
\end{displayquote}

For some physics majors, it would seem that connecting math to real life and/or being able to visualize math is part of the allure of being a physics major. Furthermore, student quotes express similar feelings regarding other courses such as real analysis, typically a very proof-heavy course. In a further example, a student comments that upper-level math courses that are taught with physical applications in mind or in a way that balances the proof-based approach with more applications can be very beneficial, in particular heaping praise on complex analysis which has many applications useful to upper-level physics and, while being proof-based, is often taught with an eye towards applications.

\begin{displayquote} \it
``So, I think that math for us really helped me understand the complex numbers that are used in physics. Similarly, as I mentioned before, many physics courses just give you this, and you have to apply them immediately after you know this fact. Sometimes you don’t fully understand the mathematical foundation of it. So, I think [complex analysis] really provided you a good foundation for understanding those complex numbers and how to use them.” (P23)
\end{displayquote}

We also found student comments suggesting that there are some cases where the content of a particular math course does not line up well with what physics students say they need, even if it is more application based. As an example, the Taylor series is a standard component of an introductory calculus series and an essential piece of a physics curriculum. However, as crucial as this is in physics some students feel that it plays more a minor role in introductory calculus. 

\begin{displayquote} \it
``... Oh, another one would be Taylor series for sure, because my [calculus course] for some reason we got behind schedule and so Taylor series we spent only week 10 on it with like the understanding that it would play a very minor role on the final and so I didn't really care much about, about Taylor series, not realizing that Taylor series would be one of the most important things I would like, return to time and time again in physics…” (P1)
\end{displayquote}

Another place where there seems to be misalignment is in quantum mechanics. A solid upper-division course on quantum mechanics involves many topics from linear algebra and most physics programs will require a course on linear algebra from the mathematics department. However, these prerequisite courses are often not taught with the goals of quantum mechanics in mind. Quantum mechanics uses complex vector spaces, Hermitian operators, and commutators; topics a typical introductory linear algebra course will not discuss. Many students expressed frustration with this mismatch and how difficult it can be to fully comprehend these topics upon encountering them for the first time. 

Further complicating things is that, according to students, some physics faculty are unaware of this misalignment. Several students noted that instructors would sometimes make comments like, “as you probably learned in your linear algebra course,” regarding advanced topics that most linear algebra courses do not cover.

Additionally, there were several student comments that suggest that a more advanced linear algebra course could be a better match with the expectations of a quantum mechanics course. 

\begin{displayquote} \it
``Obviously the number one useful thing was eigenvalues, eigenvectors, eigenfunctions and, and I feel like that was, that was maybe something that was more reinforced in [advanced linear algebra] then, and we, we kind of talked about eigenstuff in [basic linear algebra] but, but maybe in [basic linear algebra] it felt a little more arbitrary like, like I, I told my Dad that we were doing something called like, eigenvalues and he was like, oh yeah I use eigenvalues all the time in my work and, and he's, he's not a physicist, he's a totally different field but I was like, huh, well apparently these things are pretty important but I know, I had no idea how and, and my feeling is that [advanced linear algebra] reinforced that a lot more …" (P1)
\end{displayquote}

Another sub-theme was the utility of so-called ``trading zones”, an anthropological term for spaces where members of different cultures meet to trade goods and ideas that was adapted by Peter Galison~\cite{Galison_2010} to describe physicists interacting with other ``cultures", such as mathematicians and engineers, in the development of new technologies. Trading zones often develop their own languages that blend those of the cultures which interact there. Most physics departments, including ours, offer a mathematical methods course, in which mathematical topics are presented via more physical language and applications. We view such courses as trading zones that use blended math/physics language and offer an opportunity to really focus on the physical applications and visualizations of the math topics and their connection to real life systems. Furthermore, many math topics essential to physics are not discussed in the typical introductory math courses. Examples include partial differential equations (PDEs), Fourier analysis, complex analysis, and advanced linear algebra. From student comments, it seems having this space, this trading zone, to more fully explore these mathematical topics is valuable and appreciated. 

\begin{displayquote} \it
 ``... I recently took mathematical physics [math methods] last quarter and I very much enjoyed that class cuz it was a very like nice blend, I think, between, you know, physics and actual solving math stuff and I think, you know, we covered Fourier transforms, you know, maybe not as rigorously as you might in a math class but I think it was relatively satisfying. We covered Fourier transforms and complex analysis and we did use a little bit of linear algebra, you know, but you know find out eigenvalues of like a linear, linear system and I thought that was very satisfying so had I taken that earlier I think I wouldn't really have these sorts of, you know, problems I think, but on the other hand like of course you know they need to wait for people to have a certain level of maturity to take that class so I understand why it's not, you know, first quarter kind of thing.” (D4)
\end{displayquote}

As discussed, one topic rarely covered in introductory math courses that seemed problematic for our students is the Fourier transform. Even though this is typically included in a mathematical methods course it is so important that it is often introduced earlier in the physics curriculum. When this happens, it seems that many students feel that there is not enough time to get a good handle on it. 

\begin{displayquote} \it
``Yes, and honestly I still think that a math course that introduces Fourier transforms and builds them up would be good for me, and good for maybe a lot of physics majors, but yes my feeling when I first got to [quantum physics] was like we introduced it and like right after we took the second midterm, so it was hard to like feel like it would be that important just because it was like the final two weeks of the course and we were really rushing through it and, and it kind of got introduced as like, like along with the idea of like Hilbert spaces and like, and like wave functions rather than or like, like wave vectors rather than wave functions and, and, and so at that point it was like, like if it all just kind of got thrown on us feeling a bit like magic and then it was like and look there's this thing called a Fourier transform that, that fixes this or that, that solves these problems for us and, and right now you don't really need to know the specifics of it, you just need to know that it can, it can switch you between position space and, and like momentum space and, and when you hit it on a cosine you just get the, the frequency of the sine wave and it was like it was like, okay I mean this is clearly very mathy but it's just being introduced kind of as, as magic to solve the few problems that we can even apply it to, we can't even really apply it to many problems right now, we can only really apply it to." (P1) 
\end{displayquote}

\section{Major Theme 4: when do physical contexts for math help understanding?} \label{theme-4}

Physical examples can often be extremely helpful in bridging the gap between mathematics and physics. The final main theme that arose was a question of timing: when do physical contexts for math help understanding? Which is more helpful, seeing the math before, after, or at the same time as the physical context? The students had much to say about this. We even had several cases of students referencing a single class example, involving matrices, as being extremely valuable.

Many student comments promoted the idea that learning math first, in a math class, and then applying it in a physics class works well, when the class is aligned well with the needs of the physics topic, as with the introductory calculus sequence and first-year physics. There were also several comments suggesting that when this does not occur, and the math topic is introduced in a physics class right before its application, it can lead to problems. There was consensus that this way of presenting the material often doesn’t give enough time for students to fully process the new math topic, making a discussion of the physical application difficult. For example, here is a quote relating to the introduction of Lagrangians and functionals in a junior-level classical mechanics course.

\begin{displayquote} \it
``I think I wrote the other one was the Lagrangian. 'Cause then we were... it was in [classical mechanics], we were presented with the Euler-Lagrange equation, and that, that seemed... it seemed a little fast to me how it was like functionals. I didn't know that was a thing before that, and then we just have this thing like the action. Minimizing action. I was like, okay, I guess, I'll... I mean, I didn't really understand, you know, the exact derivations, and in...intuitions behind everything. But then you could still use the Lagrangian, and use all the equations, and then, you know, get the equation of motion. But then, I kind of just start with that. And I'm not sure... I wasn't exactly sure how the... how that even tied into, you know, getting those equations of motion.” (P12)
 \end{displayquote}

Furthermore, a student directly discusses this situation dealing with the introduction of Fourier transforms in a first-year quantum theory course. As mentioned, Fourier transforms are not generally presented in the typical math courses that physics majors are required to take. Here again we see a student describing simply not having enough time to digest the new mathematics enough to apply it. 

\begin{displayquote} \it
``I think the ideal way to do it would be learn it from math professor, and then learn it from a physics professor. (I: Why is that?) 'Cause, I mean, I think like things are always gonna be more complicated when you don't have, like, context around them. But at the same time, at least from my experience with both, like having at least some sort of baseline foundation of just the math part of it, and then being able to apply it." (D5)
\end{displayquote}

There also seems to be evidence that providing physical examples along with the math theory can be an effective way to teach, if students are given enough time to begin to master the material. In the upper-level complex analysis course a mixture of theory and applications seemed to work well.

\begin{displayquote} \it
“For [complex analysis] I took the class taught by Professor X, and he gave us a lot of examples in physics. I think that bridges like well from complex numbers in math to complex applications in physics. I feel like I took [complex analysis] because of my major, so I think most physics students won't take that course. I think the complex numbers are only told by that course. No other lower physics courses teach you thoroughly about complex numbers. So, I think that math for us really helped me understand the complex numbers that are used in physics. Similarly, as I mentioned before, many physics courses just give you this, and you have to apply them immediately after you know this fact. Sometimes you don’t fully understand the mathematical foundation of it. So, I think [complex analysis] really provided you a good foundation for understanding those complex numbers and how to use them.” (P23)
\end{displayquote}

It seems that that the balance and sequencing of time spent on new mathematical tools and physical applications is pedagogically important. There were several comments stating that little or no physical application was also not helpful.

\begin{displayquote} \it
Student: ``From what I... what I remember feeling was like, I remember it was just like a lot of these like counting all these row reductions and all this stuff. And I was like, `what am I doing, doing all this stuff?' Yeah, I was like, `okay, I'll do all this.' Yeah."

Interviewer: Do you remember applying it to any physical system? Or was it always just math for...?

Student: ``I think it was just math.” (P12)
\end{displayquote}

Student comments would seem to support the idea that what works best for the learning of mathematics needed for physics is a mathematical style presentation with plenty of physical applications included. It doesn't seem to be ideal when the inverse occurs, a physics presentation with mathematics topics sprinkled in. However, what seems to be the most successful approach, according to student comments, is when the math is presented first in a mathematics course and then physics uses this topic in a subsequent physics course.

Finally, well-timed and key physical applications can prove very effective. There was one key example, regarding a ring of masses and springs solved with linear algebra techniques, that really resonated with many of our students. It seemed this type of `aha’ moment could be very powerful.

\begin{displayquote} \it
``[Professor Y] really did a really good job, I thought, with the whole, the linear algebra part of it, where umm --- I don't remember how we got into it --- it was like, basically just starting with like a couple… coupled oscillators and then extending it to like, $n$ of them, like $n$ of them in a ring, and then as $n$ goes to infinity, and we were just talking about how you can write out like the force matrix or you can write out the symmetry matrix, which is equivalent, and then you can like find the eigenvalues of that. And like the way that they kind of presented eigenvalue and eigenvectors, and some of the linear algebra, I thought, was like really helpful, because it was like right in the context. They like, put it all into context, because everyone's taking [basic linear algebra]. Everyone kind of understands the ideas, but putting it in the practice was really helpful..." (D5)
\end{displayquote}

\section{Discussion} 

Our analysis of the students' surveys and interviews is summarized by our four major themes:
\begin{enumerate}
\item Students perceive physics classes as inductive, eclectic, and flexible, whereas they perceive mathematics classes as deductive, standardized, and highly structured.
\item Students were frustrated by so-called physics ``tricks" --- mathematical techniques or approximations used in physics classes that appear unclear, unjustified, unmotivated, handwavy, etc.
\item Students cite the introductory calculus sequence as being helpful for physics, and more formal, upper-division, proof-based math classes as less so. ``Trading-zone" courses, which might include upper-division applied math classes (in the math department) as well as math-methods physics classes (in the physics department), helped clear up several topics.
\item Students thought they benefit more from initial exposure to a mathematical topic in the mathematical ``style" (theorem-definition-proof), and only later applying it to a physical context after a general introduction. Learning new mathematical ideas simultaneously with new physical applications (e.g., quantum mechanics using Hilbert spaces) can easily be overwhelming.
\end{enumerate}
Our analysis of Corinne's shibboleth confirmed Redish and Kuo's claim~\cite{Redish_Kuo_2015} that physicists and undergraduates load meaning onto symbols in a way that mathematicians do not. One finding that resonated with the authors was that students were uncomfortable with ``erasing information". In our data this meant making approximations in physics (where an exact answer is not required, relevant, or possible). However, we have seen similar student reactions in math classes when manipulating inequalities in analysis: they tend to want to find the optimal bounds and are uncomfortable settling for weaker ones. 
This may be due to the fact that it is difficult to anticipate the goal of the instructors' seemingly arbitrary manipulations; students are more focused on individual manipulations rather than the overall big picture objective. It is also a general communication norm to make the most informative statements that you are able to.

Some of the results surprised us as well. First, students by-and-large were not very bothered by manipulations of infinitesimals in physics classes (for example, treating $dy/dx$ as a ratio of two small but physically meaningful quantities). In addition, the authors did not anticipate how often Fourier analysis came up as an unsettling and confusing topic. However, perhaps this has to do with the specific presentation of this topic at our university, and is not a universal experience.

Our study has several limitations. 
Although we contacted every student at our institution having a double major, or major plus minor, in mathematics and physics, our sample size was small: 20 surveys and 11 interviews. 
We only interviewed one student with a math major and physics minor. 
Our study could be repeated at a larger university, or at several, making a more quantitative analysis possible.
Although we felt that double majors would be best able to compare the cultures of mathematics and physics, one could conduct a similar study with physics majors exclusively inasmuch as they are a larger population and physics education research is particularly concerned with improving the curriculum for them. 
The students we interviewed all had senior standing at our institution and therefore might have taken many of the same classes with the same instructors. 
Some of our findings may reflect characteristics of those instructors rather than the courses or the curriculum. 
Finally, we are on the quarter system, which may imply a faster pace for covering material than semesters provide. 
A one-quarter linear algebra course taken by physics students may not reach topics relevant for quantum mechanics that could be included in a semester. 
Our first quarter of complex variables usually cannot reach the Residue Theorem. 

\subsection{Pedagogical Recommendations} 

Our study involved a small number of students at a single institution, so we should be cautious in making pedagogical recommendations based on our limited data. We venture to suggest some which might be supported by further investigation. First, and perhaps most importantly, we suggest making efforts to raise awareness of these cultural differences for both faculty and students. For example, it would be helpful for physics faculty to be more familiar with what particular mathematical topics students seem to have trouble with (e.g. Fourier transforms, Hilbert spaces.) Being more conscious of these issues could lead instructors to be more prepared and proactive when dealing with these topics, for example looking more closely at what content the typical linear algebra course at their institution covers or taking more time to introduce the Fourier transform. Furthermore, these types of conversations might reveal blind spots in the curriculum that could be fixed. For instance, knowing that the Taylor series is taught at the end of the semester at your institution could lead one to spend a little longer on the mathematical background when first discussing this topic. 

In much the same way that an instructor might explain the rationale behind new research-based teaching techniques they are using, we feel it might be beneficial to explicitly discuss these cultural differences, especially in a physics class, as the students seem to be less satisfied with mathematical explanations given by physics instructors. As our interviews indicate that students are cognizant of the differences between math and physics classes, discussion of why each discipline has developed its characteristic culture could be enlightening. These discussions could perhaps even help to start a dialogue. There seem to be times when a more mathematically inclined student is turned off by aspects of the physics teaching culture. We feel the first step in easing this tension is talking about it openly. There is even the opportunity to make light of some of these differences.

Several students suggested that it would be beneficial for physics instructors to offer extra resources, perhaps textbooks, articles, or web-based materials, for students who want more details or proofs about new or more advanced mathematical topics. We feel that this is an excellent suggestion. Perhaps the physics education community could create such resources specific to the needs of physics students. 

There also seem to be some common issues that we imagine many physics departments would share. For example, being aware that approximations and ``tricks" can prove troubling for students could lead instructors to rethink how they approach these topics, such as spending more time on why an approximation is needed/helpful or even trying to take a more mathematical approach and provide rigorous error bounds for an approximation. We were surprised by some of the resistance we found to what we felt were standard physical approximations, such as the small-angle approximation for a pendulum. We imagine that, even if math majors are more troubled by them than physics majors, most students would benefit from a more thorough analysis of some of these approximations and ``tricks". 

Finally, it's possible that physics departments might want to re-evaluate their curriculum, perhaps even surveying majors in a similar manner as we did in this study. For example, perhaps you'll find that modern quantum mechanics has reached a point where the traditional linear algebra course is no longer sufficient. Maybe there are small changes that could be made to alleviate some of the tension that arises when a new math topic is introduced. Perhaps better coordination between required sequences of math and physics courses is needed. Again, we believe just being more mindful would be a great first step. It is difficult to remember when one was a student and how foreign some of these topics may have seemed at a first encounter.

\bibliographystyle{abbrv}
\bibliography{MathPhysicsCulture.bib}

\appendix

\section{Interview Questions} \label{Interview-Questions}

We used the following working script of interview questions during our interviews:
{\small
\begin{enumerate}
\item How would you describe the role that mathematics plays within physics, in general? How would you describe the role it has played in your own physics courses? Describe instances where math has helped your understanding of physics concepts. Describe instances where it has been an obstacle to your understanding.
\item How does a physics class “feel” different from a math class? Particularly in the ways that math is presented and used?
\item What do you mean by calling a math or physics proof or problem solution “handwavey”? Can you give examples? How do you react to such proofs/solutions? [If they used this word in the survey responses]
\item Think of some examples where the same mathematics topic has been presented or used in both your math and your physics classes. Describe cases in which it was helpful for your learning to have both perspectives on that topic. Describe cases in which the different perspectives made the topic harder to learn. [This can lead to the specific subject-area questions given below, if students mention that topic.]
\item What kinds of questions (from students, or by the instructor) seem expected, encouraged, or discouraged in math and in physics classes?
\item Mathematicians have been teased for refusing to solve an equation until they can prove it has a solution, or not using an infinite series until they have proved that it converges. On the other hand, physicists have been teased for not caring whether a series converges when they make use of it. Is this consistent with your experience of math and physics professors?
\item In the survey question about the function $f(x,y) = x^2+y^2$ asking for $f(r,\theta)$ instread, would you have thought about the problem differently if it said $f(u,v) = u^2 + v^2$ instead? Would your answer for $f(r,\theta)$ have changed? Are variables like $x,y,u,v$ used differently in mathematics than in physics?
\item Can you expand more on (...)? [student-specific survey questions]
\item Please illustrate (...) with an example? [student-specific survey questions]
\end{enumerate}
}

\vspace{0.5cm}

\noindent We also drew on the following subject-specific questions:
{\small
\begin{enumerate}
\item (Fourier Transform): How were you introduced to the Fourier transform in your mathematics classes? How were you introduced to it in your physics classes? In what ways was it useful to have two different perspectives on it? In what ways was it confusing? Has the use of the F.T. in later physics or math courses improved your understanding? Has it raised further questions for you? What differences have you noticed between the math and physics versions and uses of the F.T.?
\item (Linear Algebra and Quantum Mechanics): Was your understanding of linear algebra (matrices, eigenvectors) from math classes helpful when those ideas were applied in your quantum mechanics classes? In what ways? In what ways were those aspects of quantum mechanics different from your prior mathematical understanding? How did this affect your learning? [If they limit their answers to particles moving in potentials, which involve infinite-dimensional vector spaces, suggest that they think instead about spin systems, which are closer to linear algebra as presented in math courses.]
\item (Infinitesimals): Applications of calculus in physics often involve reasoning with infinitesimal quantities ($dx, dQ, \vec{\bf ds}$). Did your math calculus courses prepare you for this? Are you comfortable with both the mathematical and physical styles of doing calculus? How do you see them as related, or different?
\item (Dirac Delta Function): Have you seen the Dirac delta function in physics, mathematics, or both? If both, how was it presented differently in each subject? In what ways do you feel that you understand it? In what ways is it confusing?
\item (Proof): What differences have you seen between the types of proof (or justification) provided in your math courses versus your physics courses? Do you feel that either (or both) styles of proof give you a solid sense that you understand why the claims are true? Or do they leave you unconvinced to some extent?
\item (Units): Another student told us that quantities with units attached (meters, grams, newtons, etc.) are very important in physics and less so in math. Have you experienced that as an important difference?
\end{enumerate}
}

\end{document}